# A Smart, Efficient, and Reliable Parking Surveillance System with Edge Artificial Intelligence on IoT Devices

Ruimin Ke, *Student Member, IEEE*, Yifan Zhuang, Ziyuan Pu, Yinhai Wang, *Senior Member, IEEE*

*Abstract*— Cloud computing has been a main-stream computing service for years. Recently, with the rapid development in urbanization, massive video surveillance data are produced at an unprecedented speed. A traditional solution to deal with the big data would require a large amount of computing and storage resources. With the advances in Internet of things (IoT), artificial intelligence, and communication technologies, edge computing offers a new solution to the problem by processing all or part of the data locally at the edge of a surveillance system. In this study, we investigate the feasibility of using edge computing for smart parking surveillance tasks, specifically, parking occupancy detection using the real-time video feed. The system processing pipeline is carefully designed with the consideration of flexibility, online surveillance, data transmission, detection accuracy, and system reliability. It enables artificial intelligence at the edge by implementing an enhanced single shot multibox detector (SSD). A few more algorithms are developed either locally at the edge of the system or on the centralized data server targeting optimal system efficiency and accuracy. Thorough field tests were conducted in the Angle Lake parking garage for three months. The experimental results are promising that the final detection method achieves over 95% accuracy in real-world scenarios with high efficiency and reliability. The proposed smart parking surveillance system is a critical component of smart cities and can be a solid foundation for future applications in intelligent transportation systems.

## I. INTRODUCTION

Urbanization has been posing great opportunities and challenges in different areas, including environment, health care, economy, housing, transportation, etc. The opportunities and challenges boost the fast advances in cyber-physical technologies and bring connected mobile devices to people's daily life. Nowadays, almost every person in the urban area is connected to the internet and has fast access to a variety of information. The convenience has been attracting more and more population to cities at an unprecedented scale and speed. In order to efficiently manage the data generated every day and use them to allocate urban resources better, the Smart City concept has been brought into people's sight. This concept combines sensors, system engineering, artificial intelligence, and information and communication technologies for the optimization of city services and operations [1], [2].

Smart City applications have a high demand for computing services to process and store big data. Cloud computing is widely recognized as the best computing service for big data processing and artificial intelligence tasks. Nevertheless, with the urban data enlarged at explosive speed, cloud computing is no more the optimal solution in many cases because it not only consumes large bandwidth but also brings latency in information transmission [3]–[5]. Meanwhile, in some extreme situations where there is a limited internet connection (speed or volume limitation), it will be challenging to process all the data on the cloud or run data processing in an online manner.

A key component of Smart City is traffic surveillance, which needs enormous computing power and storage resources to handle the city-wide surveillance video data. Recent work indicates that traffic video data dominate traffic sensing, thus generate significant data transmission, processing, and storage workload [6], [7]. However, current traffic surveillance systems are most for recording purposes (such as monitoring cameras at DOTs) [8], off-line analysis [9], and cloud computing [10]. A low-frame-rate and low-resolution video can even generate over 10Mb data per second and nearly 1Tb data per day. With the increasing deployment of city-wide traffic surveillance and growing needs in efficiency and algorithm complexity, traditional video surveillance off-line or on the cloud will not satisfy the demands shortly.

The surveillance community has been aware of the need to shift the computing workload away from the centralized cloud to the clients. Edge computing, as an answer to this, allows data generated from Internet-of-things (IoT) devices to be handled closer to the local clients where it is produced rather than transmitting it to the cloud or centralized data server for processing. Recently, researchers started to examine the availability of edge computing for traffic surveillance [1], [3],

This manuscript was first submitted for review on August 9, 2019. This work was supported in part by the Central Puget Sound Regional Transit Authority (Sound Transit) and the Pacific Northwest Transportation Consortium (PacTrans).

Ruimin Ke (email: ker27@uw.edu), Yifan Zhuang (email: zhuang93@uw.edu), Ziyuan Pu (email: ziyuanpu@uw.edu), and Yinhai Wang (yinhai@uw.edu) are all with the Department of Civil and Environmental Engineering, University of Washington.

[11]. Their studies lay a solid foundation that has excite further exploration in the field.

The two scenes that require the most traffic surveillance are roadways and parking facilities. Smart parking has been introduced to solve parking sensing and management problems in cities. A recent report shows that people spend 17 hours on average on searching for parking spaces a year, while this number for New York drivers is 107 hours [12]. To improve the parking space searching efficiency, we will require smart parking surveillance systems for automatic and online parking occupancy detection. However, it faces the same challenge as other surveillance tasks regarding the computing workload and transmission volume in the video data processing. While there are many video processing studies for smart parking surveillance [12], [13], [22]–[25], [14]–[21], exploring edge computing solutions for parking surveillance is still at an early stage. Pioneering works have investigated implementing machine learning and artificial intelligence algorithms on IoT devices [16], [25]. While they provide insightful findings to the community, their objectives are not to develop a system for real-world practice.

In this paper, we propose an edge computing surveillance system to detect parking space occupancy with smartness, efficiency, and reliability. These three metrics are defined towards the performance goals of our system: smartness is the automatic detection and pattern recognition in a parking garage scene; efficiency is about processing in a real-time and online manner; reliability means reliable and consistent detection performance in various environmental conditions. The system's processing pipeline and components are carefully designed considering data transmission volume, efficient online processing, flexibility, detection accuracy, and system robustness. Adopting the recent research on artificial intelligence and computer vision, we implement a background-based detection method and a single shot multibox detector (SSD) finetuned on a new traffic surveillance benchmark dataset on the edge devices. On the server, we improve a state-of-the-art multiple object tracking method and develop an occupancy judgement method that can handle extreme lighting conditions and occlusions. The system is first developed and set up in a lab environment, and then it is deployed in a real-world parking garage for three months. The real-world test demonstrates the system's exceptional performance in various challenging scenarios, and its potential to support a few critical future applications in smart cities.

The contributions of this work are summarized as follows.

1) This paper proposes a new system architecture with IoT and AI technologies for real-time smart parking surveillance, which splits the computation load to local IoT devices and servers targeting optimal system performance.
2) The data transmission volume is designed to be small to handle the limited network bandwidth issue in real-time video analytics.
3) A new pipeline is proposed to perform detection in extreme lighting conditions and occlusion conditions with a combination of background subtraction and SSD detection.
4) An SSD-Mobilenet detector is implemented using Tensorflow Lite on the IoT devices with transfer learning on the MIO-TCD traffic surveillance dataset.
5) A tracking algorithm is designed to operate on the server side for vehicle tracking in parking garages.
6) The thorough experimental results and findings from a variety of real-world scenarios can be a valuable reference for future research.

## II. Literature Review

From the sensing functionality perspective, recent work in the area of parking occupancy detection can be divided into three categories: wireless sensor network (WSN) solution [12], [14], [34], [26]–[33], moving sensor solution [35], [36], [45], [37]–[44], and vision-based solution [12], [13], [22]–[25], [14]–[21].

WSN solution puts one sensor node to each parking space, then multiple sensor nodes are required for the detection of multiple parking spaces. A WSN sensor should be small, sturdy, low power, and cost-effective. Over the past years, WSN sensors with different sensing abilities have been developed and deployed. The most widely used ones are magnetic, ultrasonic, infrared, and loop sensors. For example, Sifuentes et al. design a simple yet effective magnetic-based parking vehicle detection method, which incorporates a wake-up function using optical sensors [32]. Their system reliability is improved over standalone magnetic sensors. Park et al. develop an ultrasonic sensor solution for parking occupancy detection [26]. They design a multiple echo function for more accurate parking space detection than the single echo function in a real parking environment. The detection algorithms for WSN are commonly very efficient; in most cases, a thresholding method or a straightforward pipeline taking the sensor signals as input would work. However, simple algorithms lead to high false detections in certain scenarios: magnetic sensors are sensitive to large metals nearby, such as a truck in neighboring parking spaces; ultrasonic and infrared sensors can be influenced by the environment noises, such as weather and lighting conditions. Another unique feature of WSN is the large number of sensor nodes, which has high robustness to sensor failure. That is to say, even if a few sensors stop working, the system can still convey quite accurate parking information. However, this feature also leads to a high cost and scalability issue. The installation and maintenance of hundreds of sensors are inefficient, labor-intensive, or even impracticable, especially for in-ground sensors like loops.

We summarize the second category as using moving sensors for parking occupancy detection [35], [36], [45], [37]–[44]. This group of work usually uses sensors on phone apps or probe vehicles to monitor urban parking availability via crowdsensing strategies. They can support various smart parking applications in urban areas and be an alternative to static parking sensors. For example, Bock et al. conduct multiple innovative studies on using GPS sensors on the crowd of taxis to sense on-street parking space availability [40]–[42]. They start the research by answering a question of how many probe vehicles are needed for on-street parking information collection, then prove the availability and investigate more detailed aspects such as misdetection amounts and quality of sensors. Some other studies explore and test the feasibility of onboard ultrasonic sensors and camera sensors as the moving sensors for crowdsensing [36]–[38]. While recent research has demonstrated the enormous potential of crowdsensing for parking occupancy detection in the future, their applicability is still limited to specific scenarios at present. First, the cost can be very high since it requires high penetration rates of sensors (probe vehicles) to obtain sufficient parking information; second, this strategy is suitable for on-street parking detection in urban areas but not for large parking lots or rural areas where there are few moving sensors. In additional to crowdsensing, researchers have also examined single moving sensors for parking occupancy detection, such as drones [44], [45]. With the advantage of the flexibility and wide view range, drones are considered an emerging parking sensor with high cost-effectiveness.

The vision-based solution has received increasing attention for parking occupancy detection lately with the advance in computer vision and data transmission technologies [12], [13], [22]–[25], [14]–[21]. Compared to WSN, where one sensor covers a single space or moving sensors where one moving unit has one sensor, one camera sensor covers multiple spaces; thus it decreases the cost per parking space. It is also more manageable and efficient since the installation of camera systems is non-intrusive and demands no closedown of parking lots. In addition, camera is information richer than other parking sensors, which has a greater potential to support more advanced parking management. Pioneering studies model the occupancy detection as a binary classification problem on predefined regions using relatively simple features and traditional classification methods [13], [15], [17], [23], [24]. Baroffio et al. propose a method utilizing hue histogram and linear support vector machine (SVM) [23]. Their method achieves real-time processing and high accuracy on the validation data. Bulan et al. design a pipeline based on background subtraction and SVM, which has a great performance and is robust to occlusion [15]. While these traditional methods tend to have an unstable detection performance in relatively complex scenarios, they lay a great foundation for more advanced methodologies. Recently, with the emerging trend in deep learning, researchers have examined the availability of deep learning models for vision-based parking occupancy detection. For example, Nurullayev et al. propose a dilated convolutional neural network (CNN) architecture. With the specific architecture design, it is more robust and suitable for parking occupancy detection [21].

However, vision-based solutions often generate a large volume of data that may increase the cost and unreliability of data transmission. To solve this problem, vision-based systems have been implemented to edge devices instead of transmitting the original videos to the data processing center. Vitek and Melnicuk implement a histogram of gradient (HOG) based classifier on IoT devices, though the HOG feature is still handcrafted which can lead to significant errors in real-world parking scenes [25]. Some recent studies combine deep learning and IoT device to realize edge artificial intelligence to improve detection accuracy and reduce data transmission volume. Amato et al. implement CNN classifiers to determine the occupancy status of pre-defined parking spaces. Their work is an essential milestone in the area of parking occupancy detection. Though their CNNs are already quite efficient compared to most standard CNNs such as VGG [46], they still have a relatively slow classification speed even on a single image [16], [22]. Also, for this type of classification-based parking detection system, people need to manually label each parking space at local IoT devices after the installation, and in practical applications, it can be labor-intensive, not flexible, and not scalable.

This paper focuses on proposing a new vision-based solution for parking surveillance. It improves performance regarding smartness, efficiency, and reliability with specific designs on both the system architecture and the algorithms.

## III. PROPOSED SOLUTION AND DESIGN

### A. Overview

The overview of the system design is shown as a flow diagram in Figure 1. The system is composed of camera nodes, IoT devices, cellular data transmission modules, and a centralized server. In this study, the IoT devices are Raspberry Pi 3B, yet other IoT devices like Arduino and Jetson Nano could be the alternatives. The overall design considers the balance between computational load and data transmission volume, as well as the reliability and scalability of the system.

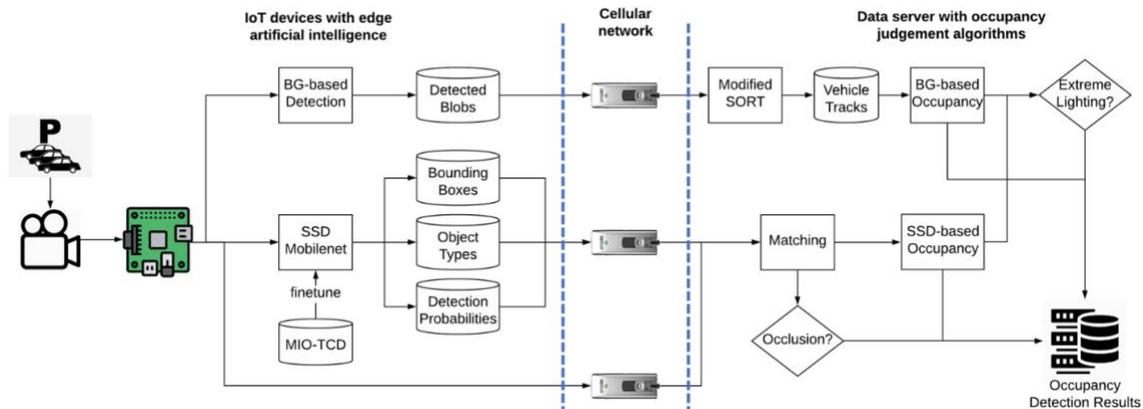

Fig. 1 Overview of the system design and methodology

Two efficient computer-vision-based object detection algorithms are implemented at the edge as two threads. They utilize the limited computation power of the IoT device to convert the raw video frames to detections in an online manner, thus largely reduce the data transmission volume and ensures efficient updates. Also, one video frame is transmitted to the data server every a few minutes for parking space labeling, results verification, and demonstration purposes.

On the server side, we propose a real-time object tracking algorithm based on SORT [47], as well as occupancy judgement algorithms considering occlusion and extreme lighting conditions. The modified SORT algorithm is implemented on the server side rather than the edge side because this design reduces the computation load at the edge while this implementation does not increase the transmission volume. Background-based occupancy detection results and SSD-based occupancy detection results are combined based on the occupancy judgement algorithms for improved robustness and accuracy.

### B. Choice and Design of the Main Pipeline

There are two major groups of pipelines in camera-based parking occupancy detection methods. In summary, in the first group, binary occupancy classifiers are developed to determine the status (occupied or vacant) of every parking space region in the camera view. The second group applies vehicle detection to localize vehicles in the whole camera view and then determines the status of parking spaces based on the matches of detection results and parking space locations. Both pipelines need a manual labeling process to mark the region of parking spaces that we are interested in. Note that automatic labeling has been attracting some research interests, but still far away from being practicable.

This labeling process has little difference between classification and detection regarding flexibility or workload in traditional server-based parking detection systems, because in either case, the labeling process is done on the server side with raw videos/images directly sent back to the server. However, in an edge computing parking system, we argue that the detection-based pipeline (the second group) is a better choice than the classification-based pipeline.

*1) The Classification-Based Pipeline and the Concern for Scalability*

First of all, please keep in mind that there are two options for the parking spaces labeling, i.e., locally on the IoT devices or on the server. For the classification-based pipeline, if the classification is done on the server, image patches of parking spaces would need to be transmitted back to the server, which significantly increases the data transmission volume and is not what we want. Hence, the classification task needs to be done on IoT devices, which means the classifier on IoT devices has to know where the parking spaces are. Thus, instead of labeling the parking spaces sitting by a server monitor, we would have to visit all IoT devices at different places, set up a monitor, look at the camera view after installation, and do the labeling. Moreover, once there is a change of the camera view (e.g., angle change or zooming in/out), someone needs to visit that IoT device again. This is not flexible or scalable. Remote connection to the IoT device could be a solution. However, in most cases, the IoT device connects to the internet using wifi or cellular network, which is not secure or friendly to remote access.

*2) The Detection-Based Pipeline and the Design*

For the detection-based pipeline, the detection has to be done at the edge. Otherwise, the system would turn into a traditional server-based system with raw videos being transmitted back to the server. As aforementioned, there is a matching stage following vehicle detection in the detection-based pipelines. In this study, we propose to move the detection to the edge side while keeping the matching stage on the server side. In this way, the system just transmits the detection results such as bounding boxes to the server for matching, rather than raw videos for detection and matching. With this design, we essentially keep the labeling process on the server side, which is flexible and scalable. To label the parking spaces, we make every edge device send one frame back to the server. This is a once-and-for-all process, and even if there is a change in the camera view, the relabeling is much less labor-intensive than the classification-based pipeline.

### C. Vehicle Detection at the Edge

There are two detection methods implemented at the edge of our system: single shot multibox detector (SSD) and background (BG) modeling detector. They work in separate threads at the edge and then their detection results are combined in occlusion or extreme lighting conditions on the server for enhanced performance.

*1) Enhanced SSD with MIO-TCD for Edge Artificial Intelligence*

SSD with a Mobilenet backbone network is the primary detector. There are different backbones for SSD, while Mobilenet has the lightest structure which makes the detection faster than other backbones. This is appropriate for an IoT device with limited computational power. We recommend using TensorFlow Lite for the SSD implementation since it is designed for deep learning on mobile and IoT devices. A normal state-of-the-art object detector like YOLO-V3 [48] with the TensorFlow platform still runs slowly with a speed lower than 0.05 frames-per-second (FPS) on Raspberry Pi 3B, and has a slightly lower detection accuracy as well. However, SSD-Mobilenet with TensorFlow Lite runs over 1 FPS on the same device according to our test. The detection results including bounding boxes, object type, and detection probabilities (how likely the result is true) are transmitted back to the server. Compared to sending videos, it reduces the data volume by thousands of times (the exact number depends on the number of detections in the video).

TensorFlow models can be converted to TensorFlow Lite models. We recommend training a TensorFlow model and then convert it to the TensorFlow Lite model. In order to improve the detection performance to make it more appropriate for practical applications, we enhance a pre-trained SSD on the Pascal VOC dataset [49] with a new traffic surveillance dataset called MIO-TCD [50], which contains 110,000 surveillance camera frames for traffic object detection training. This dataset includes a variety of challenging scenarios for traffic detection such as nighttime, truncated vehicle, low resolution, shadow, etc. To our knowledge, this is the first time MIO-TCD been adopted for parking detection, and we find it works well.

Some key parameters for the training are listed as follows: the learning rate is 0.00001, the weight decay is 0.0005, the optimizer is Adam, the batch size is 32, and the training-validation split ratio is 10:1. All layers are trainable. The training and validation loss curves, as well as some sample images at certain training steps, are displayed in Figure 2.

The enhanced SSD-Mobilenet model demonstrates great performances on traffic detection, especially in challenging surveillance image data. Figure 3 shows three examples comparing detection results between SSD trained on Pascal VOC and Pascal VOC + MIO-TCD. In the first column, the pre-trained SSD detects all big targets but misses two small targets in the back; in the second column, the pre-trained SSD misses two vehicles partially blocked by a tree; in the third column where there is snow in the nighttime, the pre-trained SSD misses most of the vehicles. Overall, the enhanced SSD produces much better detection results with few missed detections and no false detections.

*2) Background-Based Detection at the Edge*

Despite the enhanced performance of the SSD, the detection results are still not universally satisfying if your objective is to apply it to various real-world scenarios due to two reasons: (1) though much improved in speed, the SSD running 1 FPS still does not meet real-time detection at the edge, which limits the use of video temporal information; (2) deep learning model's performance depends much on the training data, but the training data can never cover all real-world scenarios, so the detector itself could still perform poorly in extreme cases. Standalone SSD-based detection may be a good option for lab demonstration, but not for field practice universally.

With this observation and consideration, we propose to add BG-based detection to the edge. BG-based detection is a widely used traditional method for traffic video surveillance that is sensitive to video noises and has no classification ability [51], [52]. But it has two advantages that can help compensate SSD: (1) it is very efficient and operates in real-time locally at the edge; (2) it has a relatively more stable detection performance in extreme scenarios where SSD does not work, though not as good in normal cases. The BG-based detection is followed with a regular blob detection step, then the bounding boxes of the detected blobs are transmitted back to the server.

### D. Data Transmission

The data transmission module in the system is composed of a 4G LTE Huawei USB Modem E397u-53, a T-Mobile data-only SIM card with 6GB monthly, and the software part. The T-Mobile data card is plugged into the 4G modem, and the modem connects with the Raspberry Pi via the USB interface. The connection of the device to the cellular network is activated via the Network Manager API in the software. The Network Manager allows automatic network connection upon start-up and automatic re-connection to the Internet whenever the connection fails. It is a reliable and helpful network connection tool that we recommend for IoT applications.

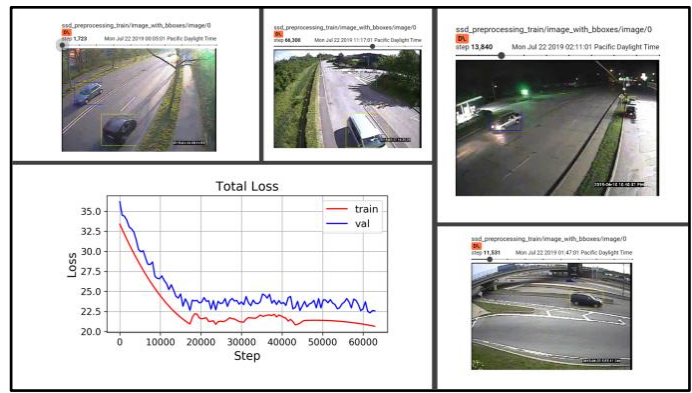

Fig. 2 The training and validation loss curves and sample images from MIO-TCD at certain training steps.

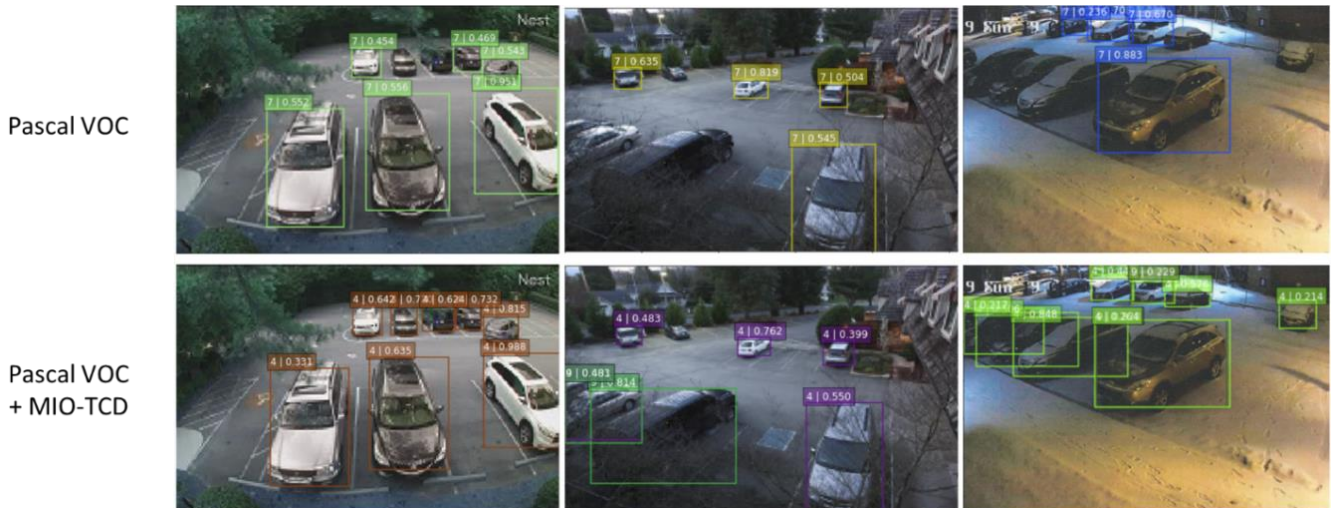

Fig. 3 The enhanced SSD-Mobilenet Detector implemented at the edge of our system has a significantly improved detection performance, especially on challenging parking scenarios in surveillance image data.

The reason we use cellular network connection for the data transmission is that the place where we do the field test does not have available wifi or ethernet. This will also be the case for many real-world IoT applications since the cellular network covers most urban areas and quite some rural areas. Other communications like Zigbee and LoRa are getting popular in IoT applications; however, they are good for short-distance communication rather than remote communication to the server. Cellular network communication is expensive with limited data amount, which, from another perspective, encourages data processing and reduction on edge. With the edge computing modules in the proposed parking system, it transmits BG-based detection results and SSD-based detection results to the server as strings. Also, the system transfers a video frame every ten minutes to the server for demonstration, validation, and space labeling. For an average camera, assuming one frame is 100Kb and the frame rate is 10 FPS (which is usually higher), and the detection results are 40Kb per minute, our system reduces the data transmission amount from around 86Gb per day per device to around 70Mb per day per device.

*E. Occupancy Judgement Pipeline and Algorithms*

With the detection results from the edge, we develop a parking occupancy judgement method on the server. This method first calculates the SSD-based occupancy based on a proposed matching algorithm and BG-based occupancy based on multiple object tracking, then combine them together considering extreme lighting conditions and occlusion conditions.

*1) SSD-Based Occupancy Detection*

The SSD-based detection results are matched with labeled parking spaces using a proposed matching algorithm. First, we design a metric for calculating the matching score of any space $i$ and detection $j$. The score $V_{ij}$ is shown below in Eq. (1),

$$V_{ij} = IoU(S_i, B_j) \times \sqrt{p_j} \qquad (1)$$

where $IoU$ is the function to calculate the intersection-over-union between two rectangles, $S_i$ and $B_j$ are the labeled parking space $i$ and the bounding box of detection $j$, and $p_j$ is the detection probability of detection $j$. Note that only detections with the category being a vehicle (e.g., car, van, bus, truck) will be kept in the detection list. Since the probability is between 0 and 1, we multiply the $IoU$ by the square root of the detection probability rather than the original probability in order to give more weight to the term $IoU(S_i, B_j)$, which should be the primary indicator of parking occupancy status than the probability.

Considering that parking occupancy status does not change very often, the status in the immediate previous time step is another indicator of the current status. Hence, a double thresholding method is adopted to filter out invalid $V_{ij}$ with two thresholds $Th_{max}$ and $Th_{min}$ ($Th_{max} > Th_{min}$). If space $i$ is occupied in the previous time step, the threshold for $V_{ij}$ will be $Th_{min}$; otherwise, if space $i$ is vacant in the previous time step, the threshold for $V_{ij}$ will be $Th_{max}$.

There are two cases that need further consideration: (1) one detection corresponding to multiple spaces and (2) one space corresponding to multiple detections. We deal with the first case first. Since one detection can only match one space at most, in the first case, the space with the largest matching score will be identified as occupied and others vacant. These should address part, if not all, of case 2. Then, if there are still case 2 for any space, its status is occupied.

*2) Modified SORT and BG-Based Occupancy Detection*

The detections from background modeling at the edge are inputs to the BG-based occupancy detection algorithm on the server. The video's temporal information is used in this module in the way of object tracking. Object tracking eliminates false detections and noises in the BG detection step and generates tracks of objects. Since our system only has the bounding boxes' location information transmitted back, the object tracking algorithm is supposed to use no more information than the boxes' locations. Tracking algorithms that require LiDAR, radar, or other image information (histogram, color, deep feature, etc.) would not work for our system [53]–[55].

A state-of-the-art tracking algorithm, called SORT [47], achieves excellent performance on efficiency and accuracy using only bounding box location information. The proposed tracking algorithm is a modified version of the algorithm. The original SORT does not have a re-identification process, which will lose track of an object if not detected for a few frames. In the BG-based detection method, only moving objects are detected. Thus, in parking lots, a vehicle is often lost with an ID switch when it stops to change direction (see Figure 4). This is also the motivation for Deep SORT, which adds a re-identification metric using deep association [53]. In our system, the Deep SORT is not possible to incorporate because it requires deep features. Hence, we add a simple yet efficient decision rule to SORT: when a new ID is assigned to an object, the algorithm searches if the new object's bounding box has enough overlap ($IoU$) with any old object within the past *m* seconds. An old object is defined as an object that was tracked in the past. If yes, the two objects are associated.

With objects' tracks and the labeled parking spaces, parking occupancy can be detected: if a track starts from inside a parking space and ends outside the space, and the tracked time of the object is over a threshold ($t\_track$ seconds), the space's status is vacant; if a track starts from outside any parking spaces and ends inside a space, and the tracked time is over a threshold ($t\_track$ seconds), this space's status is occupied.

*3) Final Detection Considering Occlusion and Extreme Lighting Condition*

The final detection considering occlusion and extreme lighting condition further improve the system accuracy in extreme cases. In the proposed system, the SSD-based method is the primary detector. The BG-based detector serves as the compensation for SSD in corner cases like occlusion and extreme lighting conditions. In normal condition, the proposed SSD-based method performs near-perfectly; however, in extreme lighting conditions such as strong fog, direct sunshine, and strong shadow, SSD or any pattern-based detector, especially when there is no following object tracking process, could have poor performance and sometimes even miss most targets. On the other hand, the BG method is relatively more stable in extreme conditions, though not as good as the enhanced SSD in normal conditions.

If extreme lighting condition warning is triggered, the two sets of results will be combined. For those spaces detected as occupied in SSD detection, their final statuses are occupied given the low false-positive rate of SSD; for spaces recognized as vacant by SSD, the system under warning will believe the BG detection results. We determine if the lighting condition is bad enough to activate the combined detection using a metric as follows,

$$r_t = \frac{ssd_t}{bg_t} + \frac{ssd_t}{ssd_{t-1}} \quad (2)$$

where $bg_t$ is the number of occupied spaces at current time *t* from the BG method, $ssd_t$ is the number of occupied spaces at current time *t* from the SSD method, and $ssd_{t-1}$ is the number of occupied spaces at the time *t-1* from the SSD method. This metric measures the difference ratio for the two detection methods and the short-time change in the SSD-based method. Extreme lighting conditions change, such as direct sunshine, usually happens in a short time and have an immediate influence on SSD. One time step here is set to five minutes based on the consideration that five minutes is short enough to ensure most space statuses are the same and long enough for a sudden lighting condition change to impact the SSD detector. The SSD-based method will be re-activated when $r_1 = \frac{ssd_t}{bg_t}$ returns from very small to close to 1.

Occlusion is often caused by a large vehicle's appearance and camera angle. In this study, we consider the case that one vehicle blocking two spaces, while other types of occlusion are even rarer. The system checks routinely if a bounding box of a vehicle covers two spaces. Here "cover" means two adjacent parking spaces are both at least *occ*% inside a vehicle's bounding box. In the occlusion case, the system first determines which space this vehicle is in by comparing the center of the two spaces, and the one closer to the camera (closer to the image bottom) is the space the vehicle in. For the other space, the system will use the BG-based results when it is occluded, because the object tracking will still give a clue which spaces a vehicle starts from or ends in.

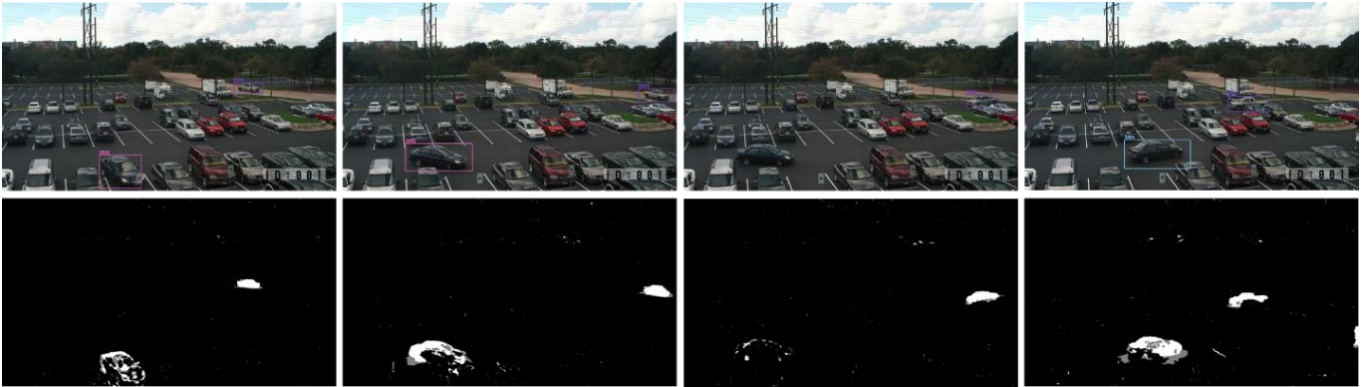

Fig. 4 BG-based detection and the original SORT tracking results. In a parking lot scene, the car at the bottom-left is lost with an ID switch due to its stop to change direction. Our modified SORT algorithm on the server solves this problem, thus reduce the error in parking occupancy detection.

## IV. EXPERIMENTAL RESULTS AND ANALYSIS

### A. Preliminary Test and Parameter Setting

This study was sponsored by Sound Transit, which is a public transit agency serving the Seattle metropolitan area in the U.S. state of Washington. The preliminary test was conducted at the Smart Transportation Applications and Research Laboratory (STAR Lab) of the University of Washington. Before the field test, over five hours of parking lot surveillance video clips and over two thousand parking lot images were collected from the internet and the Angle Lake parking garage for preliminary test and system parameter setting. Note that the Angle Lake parking garage is the test site of this study, which is a busy parking garage located near the Sea-Tac International Airport with 1,160 available parking spaces. The parameter setting is critical to the operation of the system. Table I summaries the key parameters of the system and their setting for the field test in the study based on the preliminary lab test.

There are seven parameters that need to be set. The given parameter values in Table I can be a reference for the general parking context. For some specific cases, these parameters may need to be adjusted for optimal system performance. Th$_{max}$ is always set larger than Th$_{min}$ according to their definition. A general rule for setting these two parameters is that if neighboring parking spaces have larger overlaps from the camera angle, the two parameters may need to be set larger to avoid false matching. We do not suggest modifying IoU_track or t_track in most cases. They are parameters for the tracking algorithm, which are not sensitive to the context. But they can also be adjusted based on the drivers' behaviors in a region or the user preference. The r$_t$ and r$_1$ can be adjusted based on the number of parking spaces covered in the camera view and the weather conditions in a region. If a camera view includes quite a few parking spaces, say, more than 6, we do not recommend changing r$_t$ and r$_1$ by much. But if a camera covers only 2 or 3 spaces, extra efforts would be expected in the parameter setting because you may not be able to tell whether a missed detection is due to sudden lighting changes or other factors. The difference in weather conditions of different regions could also influence the setting of these two parameters. The last parameter $occ\%$ can be adjusted based on camera angles; the more the overlap of neighboring parking spaces, the larger $occ\%$ should be.

TABLE I DESCRIPTION AND SETTING OF KEY SYSTEM PARAMETERS

| Parameter | Parameter Description | Default Value |
|---|---|---|
| $Th_{max}$ | The larger threshold for the matching score $V_{ij}$ in the SSD-based occupancy detection | 0.25 |
| $Th_{min}$ | The smaller threshold for the matching score $V_{ij}$ in the SSD-based occupancy detection | 0.1 |
| IoU_track | The intersection-over-union threshold to determine if a new object is associated with an old object in the modified SORT algorithm | 0.6 |
| t_track | The time threshold in the unit of seconds to determine if a new object is associated with an old object in the modified SORT algorithm | 8 |
| $r_t$ | The threshold to trigger the bad environment lighting warning to the system | 0.8 |
| $r_1$ | The threshold to re-activate the normal detection pipeline in the system | 0.7 |
| $occ\%$ | The portion of a space inside a vehicle bounding box to determine if the space is covered by the vehicle in camera view in the occlusion judgement | 90% |

### B. System Installation and Data Collection

Two IoT devices were installed, one on the sixth floor and another on the third floor of the Angle Lake parking garage. Figure 5 shows the installation of the IoT device on the sixth floor, the data server set up at the STAR Lab, and camera views from the two cameras. The sixth floor was an outdoor parking scene and the third floor was indoor. In the field test, our cameras monitored sixteen parking spaces, which were No.1013 – 1022 on the sixth floor and No.503 – 508 on the third floor. The purpose of choosing the sixth floor was to test the system performance outdoor, particularly how it performed in different weather, temperature, and time of day. The third floor was selected to test the indoor performance. The low ceiling height of this floor and the installation angle of the camera created challenges such as occlusion, which was meaningful for testing the system. The system was operating for three months from September 16, 2018, to December 15, 2018, at the Angle Lake parking garage. In total, only less than 20Gb data was transmitted back onto the STAR Lab server from the two IoT devices in three months.

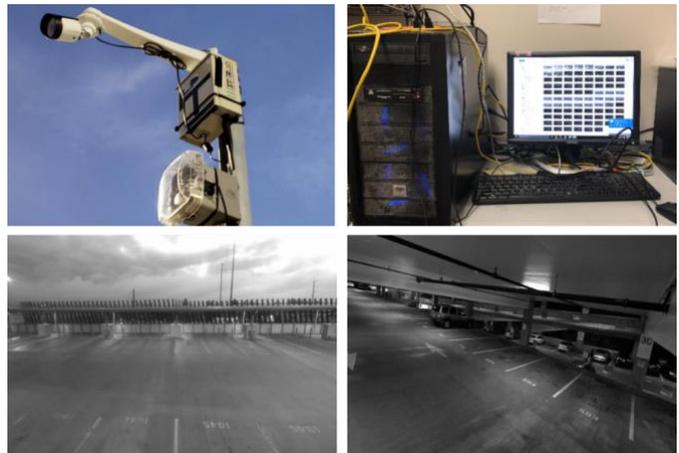

Fig. 5 System installation at the Angle Lake parking garage (top-left) and the server set up at STAR Lab (top-right); the bottom row displays the camera views of the devices we installed on the sixth floor (bottom-left) and the third floor (bottom-right).

### C. Results and System Evaluation

Every ten minutes, a video frame was transmitted to the server for validation and demonstration. Overall, the system achieves 95.6% detection accuracy during the three months. Figure 6 presents the sample detection results and Table II shows the summarized statistics of the experiment. In this table, we divided the detection conditions into multiple categories based on the weather, time (day or night), day (weekday or weekends), and floor (indoor or outdoor). The categorization was done by manually classifying the scene pictures captured by the cameras as well as checking the historical weather records. And the detection performance for each category was summarized using accuracy as the metric ( $accuracy = \frac{correct\ detection}{total\ number\ of\ spaces} \times 100\%$ ).

*1) Cloudy and Rainy Scenarios*

The weather was classified into four categories, which are sunny, rainy, cloudy, and foggy conditions. For the third floor (indoor), there were no significant accuracy differences across the four weather conditions, while the sixth floor (outdoor) was influenced more by the weather. Among the four weather conditions, the system performed the best in cloudy conditions, reaching 97.5% accuracy on weekdays and 99.2% on weekends, due to the relatively consistent lighting conditions over the video field of views (Figure 6(a)). The second highest was in rainy conditions with an accuracy of 93.7% on weekdays and 96.2% on weekends, where the lighting conditions were similar to cloudy days. However, raindrops on the lens might sometimes block parking spaces though a camera shelter was employed to protect the camera (see Figure 6(b)). This was rare but the main cause of its lower accuracy than cloudy days. In rainy and cloudy days, background-based occupancy detection was seldom activated.

*2) Sunny and Foggy Scenarios*

The detection accuracies in sunny conditions and foggy conditions were both lower than cloudy and rainy days. We carefully examined the ground-truth images and found out the reasons. In sunny conditions, there were sometimes strong shadows of the vehicles, and reflections towards the camera on the sixth floor. In our case, shadows and reflections could significantly change the visual appearances of the vehicles, thereby confusing the feature extraction processes, especially when the occupancy was high: since vehicles were very close to each other and the ten parking spaces were covered by just one camera from around 25 – 30 feet away, the shadow of the fence on the sixth floor and the reflection sometimes could influence multiple vehicles; also, the shadow of one vehicle could not only change the visual appearance of itself but also the vehicles next to it. Foggy conditions had the lowest detection accuracy for the outdoor parking with 85.7% on weekdays and 91.6% on weekends. Our observation indicated that the thicker the fog was, the lower the detection accuracy was. In sunny and foggy conditions, background-based occupancy detection was activated more than rainy and cloudy conditions. They help compensate the SSD detection and significantly improve the accuracy following the proposed extreme condition detection pipeline. Though the overall accuracies of sunny and foggy days were still lower than the average, it was already increased a lot over standalone SSD-based occupancy detection. For example, in Figure 6(c) and (d), it can be seen in the extreme lighting conditions like direct sunshine and strong fog, even the enhanced SSD's performance significantly decreased with higher missed rates and lower detection probabilities.

*3) Nighttime, Weekend, and Occlusion*

It was interesting to note that the overall detection accuracy at night was higher than that during the day, and the detection performed better on weekends than on weekdays (see Figure 6(e)(f)(g)). These results were mainly caused by the property of the SSD detector. Our detector was finetuned to have a very low false-positive rate in order to achieve high precision. In other words, in case there was a false detection, it is more likely that an occupied space was recognized as a vacant space, rather than a vacant space being recognized as occupied. According to this fact, it was interpretable that the accuracy at night and weekend were overall higher than in the day and weekday, because the traffic volume and the number of occupied spaces were lower at night and over the weekend. Moreover, the enhanced SSD got good detection results in dark due to the large number of image training samples taken at night in the MIO-TCD dataset.

It was observed that the detection performance was more consistent on the third floor with a lower variance than the sixth floor in different weather conditions. Also, the main causes of errors in detection for the two floors were actually different. It was found that most errors on the sixth floor were caused by extreme lighting conditions (shadow, reflection, and fog), while on the third floor the errors were caused more by occlusion. For spaces 503, 504, and 505, the detection accuracy was almost 100% for all scenarios. However, due to the installation angle of the camera and the low ceiling height, spaces 506, 507, 508 could be partially or fully blocked by the vehicles parking next to them. Figure 6(h) showed an example of a van parking in space 507 completely blocking space 508, but the occlusion case was handled by our proposed pipeline. Note that occlusion was dealt with by our system with (1) SSD on partially blocked vehicles, or (2) BG-based detection and tracking if an occlusion warning was triggered.

TABLE II SYSTEM DETECTION ACCURACY STATISTICS

|  | Sunny | Rainy | Cloudy | Foggy | Day | Night | Average |
|---|---|---|---|---|---|---|---|
| Average | 91.4% | 93.5% | 95.5% | 89.9% | 92.7% | 98.4% | **95.6%** |
| On third Floor (Weekday) | 92.3% | 91.8% | 92.6% | 92.0% | 92.2% | 99.1% | 95.7% |
| On third Floor (Weekend) | 94.3% | 94.5% | 93.9% | 93.1% | 94.0% | 99.0% | 96.5% |
| On sixth Floor (Weekday) | 88.5% | 93.7% | 97.5% | 85.7% | 91.7% | 97.3% | 94.5% |
| On sixth Floor (Weekend) | 93.8% | 96.2% | 99.2% | 91.6% | 95.4% | 98.9% | 97.2% |

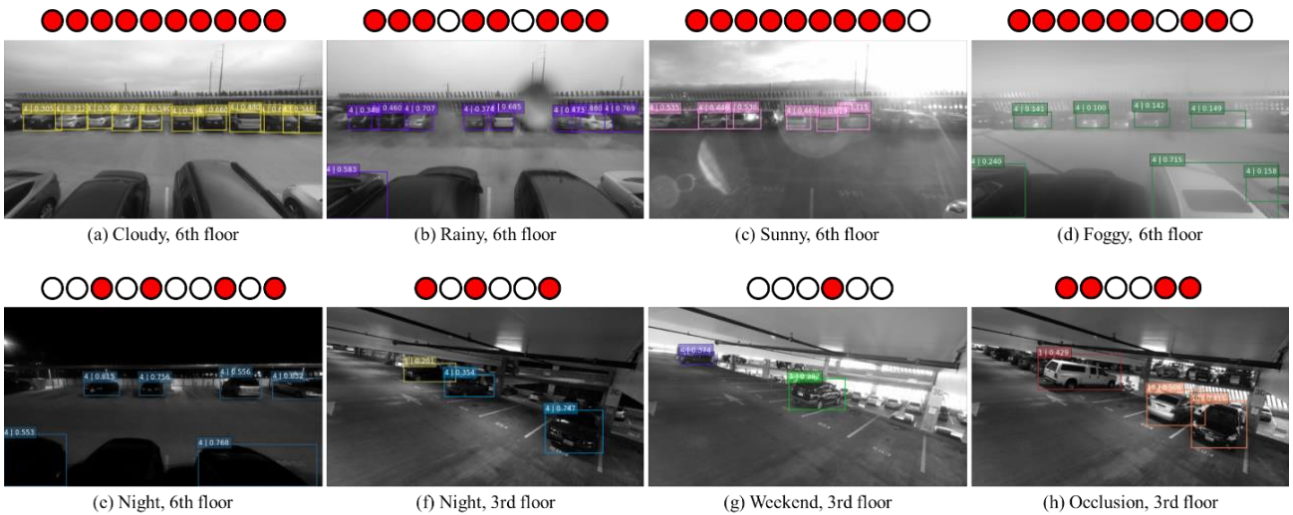

Fig. 6 The figure displays final detection results (on the top of each image) and enhanced SSD-based detection results (in images) in representative scenarios. Extreme lighting condition or occlusion warning was triggered in scenario (c) (d) and (h), in which BG-based detection was activated.

## D. Comparison with State of the Arts

We summarized the comparison between this study and the state of the arts in automatic parking surveillance in Table III. First of all, the system inputs are different among different systems. Taking images as the input could be straightforward but will lose the temporal information. Systems that have IoT devices leverage the power for efficient transmission and onboard processing so that they enable real-time occupancy updates. Compared to [15] and [19] which also use IoT devices, this study deploys the computations on both the edge and the server, which helps efficiently handle the workload; also, we implement a detection-based pipeline thereby does not need labeling on every IoT device. Please note that though previous systems including IoT devices do not have computation tasks on the server, they do need servers as part of the system for data storage. Regarding primary algorithms, this study and [20] are among the first efforts to use the deep-learning-based object detectors for parking (Faster R-CNN and SSD). However, SSD is the latest one-stage object detector, which is faster than the two-stage Faster R-CNN. Thus, it can better support edge computing. Our study also has the largest number of frame samples for training (127,125 frames). The work [19] has 390,000 image patches for training, where one patch is one parking space they cropped from the original frame. They do not mention how many frames they use. This study and [15] are the two using real-world data for validation. While others using a few frames or images. Our validation covers a relatively long time (three months) and more scenarios. In terms of system efficiency, [18]–[20] do not mention their processing speeds or efficiency measures in their papers since that is not their main focus. [15] achieves 5 frames per second processing on their desktop with no artificial intelligence methods, but is still an impressive performance in 2013. [22] mentions their CNN can process 50 spaces in an image per 15 seconds (about 3 spaces per second). The proposed system achieves about 1 frame per second, which is faster than most existing systems (some not shown in the table). Updating a parking lot's occupancy status every one second is sufficient in most cases. The state of the arts all achieve great accuracy (over 90%). Due to the different inputs and designs of these systems and the lack of a widely accepted public parking video dataset (image datasets does not work for many systems), the system accuracy of each system cannot be directly compared at this time.

## E. System and Data Applicability

Figure 7 shows an example of the occupancy data, which was automatically collected on the week Nov 12 – Nov 18, 2018. The plots give an intuition on the parking occupancy patterns in the garage. The proposed system and the real-time parking occupancy data generated by the system can be valuable resources to support a variety of intelligent transportation applications, such as smart parking management, advanced infrastructure systems, and connected and automated vehicles.

TABLE III COMPARISON BETWEEN THE PROPOSED SYSTEM AND STATE OF THE ARTS

| Research work | Bulan et al. 2013 [15] | Ling et al. 2017 [18] | Amato et al. 2017 [22] | Cho et al. 2018 [19] | Nieto et al. 2018 [20] | **This study** |
|---|---|---|---|---|---|---|
| System input | Video | Video | Image | Image | Multiple videos | **Video** |
| Computation platform | Desktop | IoT devices | IoT devices | NA | Desktop | **IoT devices and server** |
| Process mode | Post analysis | Onboard processing | Onboard processing | Post analysis | Post analysis | **Onboard processing** |
| Pipeline logic | Detection | Classification | Classification | Classification | Detection | **Detection** |
| Primary algorithms | SVM, HOG, BG | Haar, F-test | CNN | Random forest | Faster R-CNN, fusion | **SSD, BG, SORT, fusion** |
| # of training frames | 1,800 | 469 | 4,323 | 390,000 (patches) | 23,741 | **127,125** |
| Validation data | Several days real-world validation | 90 detections | CNRPark + EXT image dataset | 24,000 image patches | 1,000 frames | **Three months real-world validation** |
| Testing scenarios | Outdoor, sunny, cloudy, rainy, daytime, occlusion | Outdoor, daytime | Outdoor, sunny, cloudy, rainy, daytime | Indoor | Outdoor, clear, rainy, daytime, nighttime | **Outdoor, indoor, occlusion, sunny, cloudy, rainy, foggy, daytime, nighttime** |
| System efficiency | 5 frames per second | NA | 3 spaces per second | NA | NA | **1 frame per second** |
| System accuracy | 93.9% | 91% | > 90% | 98.6% | > 90% | **95.6%** |

*1) Application in Smart Parking Management*

One of the key components in intelligent transportation systems is smart parking management. Smart parking management targets improving the efficiency in parking resource allocation, parking information dissemination, and parking space searching with high accuracy, robustness, and low cost. Smart parking management also has the ability or potential to help mitigate traffic congestion and other societal problems. Parking prediction, dynamic parking pricing, real-time parking guidance, etc., most smart parking management strategies and functions need parking occupancy data as the input. Thus, the proposed system, which is designed to work in a wide range of scenarios with low cost and high performance, provides foundations to smart parking applications as well as modern transportation management.

*2) Application in Advanced Infrastructure Systems*

Future advanced infrastructure systems will solve problems related to buildings, bridges, pipelines, roadways etc. by combining conventional physical assets with emerging cyber technologies in computer science, system engineering, and other fields. Edge computing will be a critical component of infrastructure management of tomorrow, especially with the emergence of 5G communication. A cost-effective, real-time, reliable, and scalable edge computing system for parking occupancy detection will offer new solutions and opportunities to smart city developments by making infrastructures like buildings and roadways smarter, more efficient, and more sustainable.

*3) Application in Connected and Automated Vehicles*

Parking occupancy data will be an essential component in connected and automated vehicle applications. First of all, automated vehicles will need to find parking themselves, which will be completed faster with their own systems communicating with nearby parking facilities. Additionally, parking facilities will serve as crucial nodes in a traffic roadway network to support V2I functions. IoT devices monitoring parking spaces will indirectly obtain the traffic conditions nearby from parking occupancy, thus help network-wide decision making in the era of connected and automated vehicles.

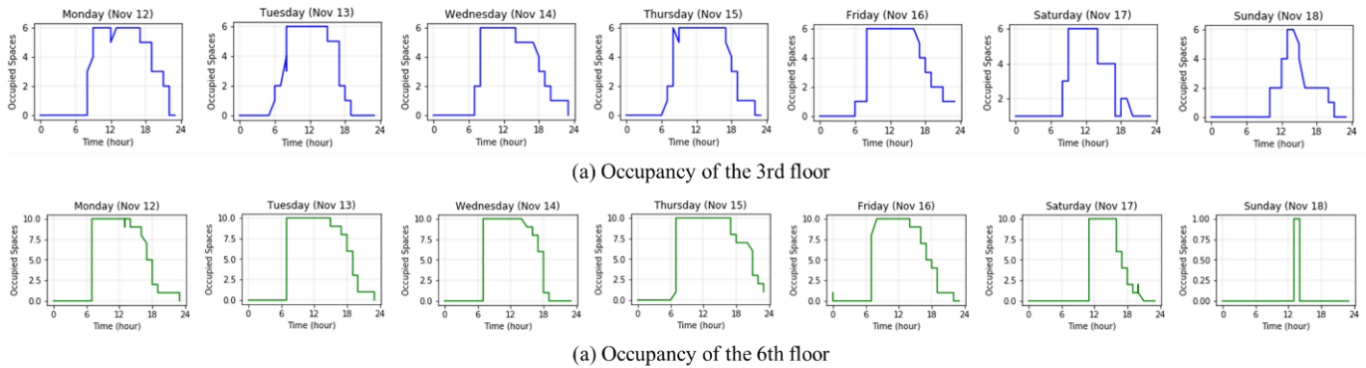

Fig. 7 The parking occupancy patterns of the week Nov 12 – Nov 18, 2018.

## V. Conclusion

In this paper, we described the design, development, deployment, and evaluation of a smart, efficient, and reliable parking surveillance system with edge artificial intelligence on IoT devices. We did a thorough literature review on smart parking surveillance. The proposed system was among the first efforts in applying edge computing techniques to real-world parking surveillance. The system processing pipelines and algorithms were carefully designed for the purpose of reasonably shifting computing workload to the edge, thus significantly reduce data transmission volume and enable efficient online parking occupancy detection. Experiments were conducted first in the STAR Lab and then in the Angle Lake parking garage for three months. The system components collaborated very well under the proposed scheme. The system achieved 95.6% overall detection accuracy in different scenarios including indoor, outdoor, cloudy, rainy, sunny, foggy, occlusion, daytime, and nighttime situations. The design has multiple advantages over the state-of-the-art parking surveillance systems and has a bright prospect in the applications of smart city and intelligent transportation systems.

## Acknowledgment

The authors would like to express our gratitude to the Central Puget Sound Regional Transit Authority (Sound Transit) and the Pacific Northwest Transportation Consortium (PacTrans) for funding this research, and the help Sound Transit provided in the system installation. We would also like to thank the reviewers for volunteering their time to review this manuscript.

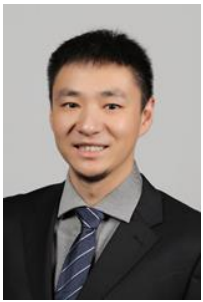

**Ruimin Ke** (S'15) received his B.E. degree from the Department of Automation at Tsinghua University in 2014, and the M.S. degree from the Civil and Environmental Engineering at University of Washington in 2016. He is currently working toward the Ph.D. degree in Civil and Environmental Engineering at University of Washington. Since 2014, he has been a Research Assistant with the Smart Transportation Applications and Research Laboratory (STAR Lab) at University of Washington. His research interests include intelligent transportation systems, transportation data science, smart city, autonomous driving, internet of things, and computer vision. Mr. Ke received the 2018 Outstanding Graduate Student Award presented by ITE Washington and the 2019 Michael Kyte Outstanding Student of the Year Award presented by PacTrans, USDOT Region 10 University Transportation Center.

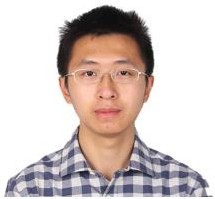

**Yifan Zhuang** received his B.E. degree in Automation from Tsinghua University, Beijing (2016) and a M.S. degree in Civil Engineering from University of Washington, Seattle (2019). Currently, he is working toward the Ph.D. degree in Civil Engineering at University of Washington. Since 2016, he is working as a Research Assistant at Smart Transportation Application and Research (STAR) Lab, University of Washington. His major research interests are application of edge computing in transportation field, transportation sensing technologies, and computer vision.

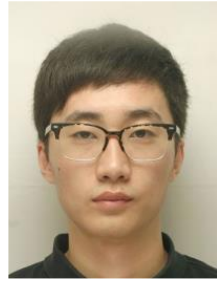

**Ziyuan Pu** received a B.S. degree in transportation engineering from Southeast University, Nanjing, China (2010) and a M.S. degree in civil and environmental engineering from University of Washington, Seattle, US (2015). He is currently working toward the Ph.D. degree in civil and environmental engineering at University of Washington, Seattle, WA, US. Since 2015, he has been a Research Assistant with the Smart Transportation Applications and Research Laboratory (STAR Lab), University of Washington. His research interests include remote sensing technology, deep learning, machine learning, traffic data mining and intelligent transportation systems.

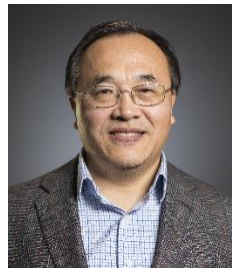

**Yinhai Wang** (SM'18) is a professor in transportation engineering and the founding director of the Smart Transportation Applications and Research Laboratory (STAR Lab) at the University of Washington (UW). He also serves as director for Pacific Northwest Transportation Consortium (PacTrans), USDOT University Transportation Center for Federal Region 10. He has a Ph.D. in transportation engineering from the University of Tokyo (1998) and a master's degree in computer science from the UW. Dr. Wang's active research fields include traffic sensing, urban mobility, e-science of transportation, transportation safety, etc. He has published over 190 peer reviewed journal articles and delivered more than 180 invited talks and nearly 300 other academic presentations.

Dr. Wang serves as a member of the Artificial Intelligence and Advanced Computing Committee of the Transportation Research Board. He is a fellow with American Society of Civil Engineers (ASCE) and past president of the ASCE Transportation & Development Institute (T&DI). He is a member of the IEEE Smart Cities Technical Activities Committee and was an elected member of the Board of Governors for the IEEE ITS Society from 2010 to 2013. He co-chaired the First and Third IEEE International Smart Cities Conferences. Additionally, Dr. Wang is associate editor for three journals: Journal of ITS, PLOS One, and Journal of Transportation Engineering. He was winner of the ASCE Journal of Transportation Engineering Best Paper Award for 2003 and the Institute of Transportation Engineers (ITE) Innovation in Education Award for 2018.